\documentclass[%
 reprint,
 amsmath,amssymb,a
 aps,
]{revtex4-2} 

\usepackage{graphicx}
\usepackage{dcolumn}
\usepackage{bm}
\usepackage[colorlinks,linkcolor=blue,citecolor=blue]{hyperref}
\usepackage{graphicx}

\usepackage{url}
\usepackage{ulem}
\usepackage{amsmath}
\usepackage{amsfonts}
\usepackage{textcomp}
\usepackage{subfigure}
\usepackage{verbatim}
\usepackage{rotating}

\usepackage{multirow} 
\usepackage{booktabs}
\usepackage{lineno}

\def \gev  {~\mbox{GeV}}

\def \gevcc{~\mbox{GeV/$c^2$}}

\begin{document}

\title{Search for the massless dark photon with $D^0\to\omega\gamma'$ and $D^0\to\gamma\gamma'$}

\author{
M.~Ablikim$^{1}$, M.~N.~Achasov$^{4,c}$, P.~Adlarson$^{76}$, O.~Afedulidis$^{3}$, X.~C.~Ai$^{81}$, R.~Aliberti$^{35}$, A.~Amoroso$^{75A,75C}$, Y.~Bai$^{57}$, O.~Bakina$^{36}$, I.~Balossino$^{29A}$
% ...
\\
\vspace{0.2cm}
(BESIII Collaboration)\\
\vspace{0.2cm} {\it
$^{1}$ Institute of High Energy Physics, Beijing 100049, People's Republic of China\\
% ...
$^{3}$ Bochum  Ruhr-University, D-44780 Bochum, Germany\\
$^{4}$ Budker Institute of Nuclear Physics SB RAS (BINP), Novosibirsk 630090, Russia\\
% ...
$^{29}$ INFN Sezione di Ferrara, (A)INFN Sezione di Ferrara, I-44122, Ferrara, Italy; (B)University of Ferrara,  I-44122, Ferrara, Italy\\
% ...
$^{35}$ Johannes Gutenberg University of Mainz, Johann-Joachim-Becher-Weg 45, D-55099 Mainz, Germany\\
$^{36}$ Joint Institute for Nuclear Research, 141980 Dubna, Moscow region, Russia\\
% ...
$^{57}$ Southeast University, Nanjing 211100, People's Republic of China\\
% ...
$^{75}$ University of Turin and INFN, (A)University of Turin, I-10125, Turin, Italy; (B)University of Eastern Piedmont, I-15121, Alessandria, Italy; (C)INFN, I-10125, Turin, Italy\\
$^{76}$ Uppsala University, Box 516, SE-75120 Uppsala, Sweden\\
% ...
$^{81}$ Zhengzhou University, Zhengzhou 450001, People's Republic of China\\
\vspace{0.2cm}
% ...
$^{c}$ Also at the Novosibirsk State University, Novosibirsk, 630090, Russia\\
% ...
}

}


\begin{abstract}
Using $7.9~\rm{fb^{-1}}$ of $e^+e^-$ collision data collected at $\sqrt{s}=3.773$ GeV with the BESIII detector at the BEPCII collider, we search for the massless dark photon with the flavor-changing neutral current processes $D^0\to\omega\gamma'$ and $D^0\to\gamma\gamma'$ for the first time.
No significant signals are observed, and the upper limits at the 90\% confidence level on the massless dark photon branching fraction are set to be $1.1\times10^{-5}$ and $2.0\times10^{-6}$ for $D^0\to\omega\gamma'$ and $D^0\to\gamma\gamma'$, respectively. These results provide the most stringent constraint on the new physics energy scale associated with $cu\gamma'$ coupling in the world, with the new physics energy scale related parameter $|\mathbb{C}|^2+|\mathbb{C}_5|^2<8.2\times10^{-17}\gev^{-2}$ at the 90\% confidence level. 
\end{abstract}

  

\oddsidemargin  -0.2cm
\evensidemargin -0.2cm
\maketitle

Although the standard model (SM) has achieved great success in high-energy physics, some questions like, e.g., dark matter, matter and anti-matter asymmetry, fermion mass hierarchy remain unresolved. So-called ``dark" sectors have been theorized, named as such due to their assumed extremely faint interactions with the visible sector.
The dark photon $\gamma'$, which is introduced in the minimum extension of the SM with an additional Abelian gauge group~\cite{Holdom:1985ag,He:2017zzr,Chiang:2016cyf}, serves as a portal between the SM matter and dark sector.
There are two distinct categories of dark photons: the massive dark photon, which arises when the symmetry of the additional Abelian gauge group is spontaneously broken, and the massless dark photon with the symmetry remaining unbroken~\cite{Dobrescu:2004wz,Pan:2018dmu,Fabbrichesi:2017vma,Su:2020xwt,Su:2019ipw,Li:2024iqv,Su:2020yze,Gabrielli:2016cut,Biswas:2022tcw,Fabbrichesi:2020wbt}.

The massless dark photon has garnered significant attention in addressing SM anomalies.
In astrophysics and cosmology, dark matter potentially interacts via a non-gravitational long-range force mediated by the massless photon, offering a potential explanation for galaxy formation and dynamics~\cite{Gradwohl:1992ue,Carlson:1992fn,Foot:2004pa,Ackerman:2008kmp,Fan:2013tia,Foot:2014uba,Heikinheimo:2015kra,Agrawal:2016quu,Acuna:2020ccz}.
In Higgs physics, the massless dark photon is introduced to generate the Higgs Yukawa couplings from the dark sector, presenting a natural solution to the SM fermion mass hierarchy puzzle~\cite{Gabrielli:2013jka,Gabrielli:2016vbb}, the origin of the CKM matrix structure~\cite{Gabrielli:2019sjg}, and the vacuum instability problem in the SM Higgs sector~\cite{Gabrielli:2021lkw}.
In flavor physics, interestingly, BELLE-II recently reported the evidence of $B^+\to K^+\nu\bar{\nu}$ decay exceeding the SM expectation~\cite{Belle-II:2023esi}.
One potential explanation for this anomaly is the involvement of the massless dark photon as an off-shell mediator~\cite{Gabrielli:2024wys}.

Experimentally, extensive searches have imposed stringent constraints for the massive dark photon~\cite{BaBar:2014zli,BaBar:2017tiz,KLOE-2:2018kqf,NA482:2015wmo,NA62:2019meo,NA64:2023wbi,LHCb:2019vmc,BESIII:2017fwv,BESIII:2018aao,BESIII:2022oww,BESIII:2018qzg,Cline:2024qzv}, while the massless dark photon remains significantly less constrained.
The two categories of dark photons are distinct, and the constraint on the massive type cannot be applied to the massless one.
To date, there have been only three direct experimental measurements conducted for the massless dark photon: $H\to\gamma\gamma'$~\cite{CMS:2020krr,ATLAS:2022xlo,ATLAS:2024cju}, $\mu\to e\gamma'$~\cite{TWIST:2014ymv} and $\Lambda^+_c\to p \gamma'$~\cite{BESIII:2022vrr}, with no significant signals observed. The sensitivity of these searches can not reach the theoretically predicted allowed region for the massless dark photon~\cite{Gabrielli:2016cut,Biswas:2022tcw}.
Therefore, more stringent measurements are required to address the experimental gaps of the massless dark photon, a non-negligible aspect within the dark photon framework in the dark sector.

Unlike the massive dark photon, which can interact with SM particles through a renormalizable dimension-four operator, the massless dark photon can only couple to SM particles through operators of higher dimension~\cite{Fabbrichesi:2020wbt}.
A dimension-six operator has been proposed to provide a connection between SM fermions and the massless dark photon~\cite{Dobrescu:2004wz}:
\begin{eqnarray}
\begin{aligned}
\mathcal{L}_{\rm{NP}}=&\frac{1}{\Lambda^2_{\rm{NP}}} ( C^U_{jk} \bar{q}_j \sigma^{\mu\nu} u_k \tilde{H} + C^D_{jk} \bar{q}_j \sigma^{\mu\nu} d_k H \\
&+ C^L_{jk} \bar{l}_j \sigma^{\mu\nu} e_k H + h.c. ) F'_{\mu\nu},
\end{aligned}
\label{eq:dimension-six operator}
\end{eqnarray}
where $\Lambda_{\rm{NP}}$ is the effective mass, indicating the new physics (NP) energy scale, $C^U_{jk}$, $C^D_{jk}$, and $C^L_{jk}$ are the up-type, down-type, and charged-lepton-type dimensionless coefficients, respectively, depending on the NP and not necessarily related to one another, $j(k)=1,2,3$ is the generation tag of the SM particle. More details can be found in Ref.~\cite{Dobrescu:2004wz}.
The first three terms in this equations are the couplings between the massless dark photon and the up-type quarks, down-type quarks, and charged leptons, where the flavors of the two quarks or leptons could be identical or different, differing to flavor diagonal of the tree-level couplings of the massive dark photon.

\begin{figure}[htbp] \centering
	\setlength{\abovecaptionskip}{-1pt}
	\setlength{\belowcaptionskip}{10pt}
 
        \subfigure[]
        {\includegraphics[width=0.238\textwidth]{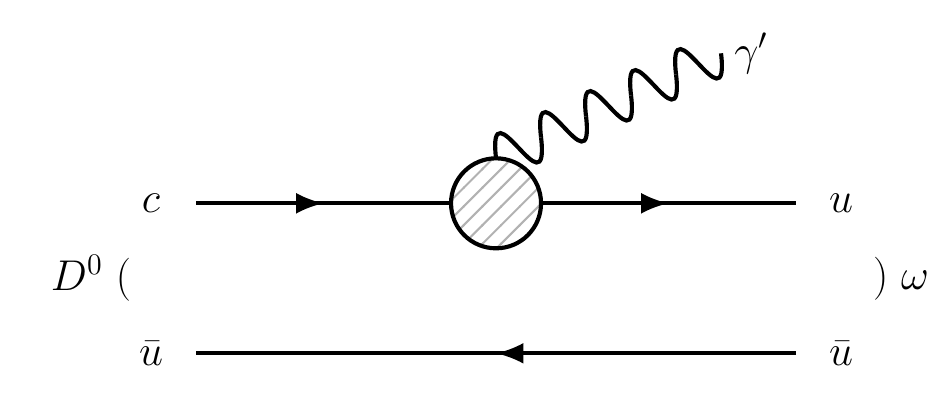}}
        \subfigure[]
        {\includegraphics[width=0.238\textwidth]{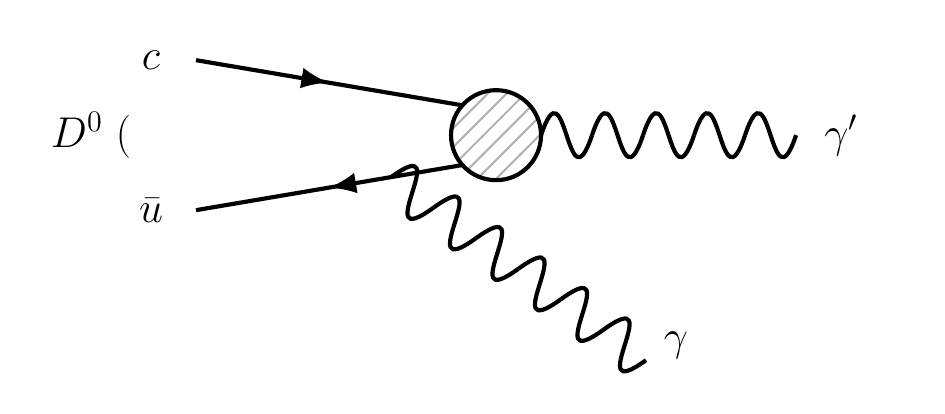}}\\

        
	\caption{The Feynman diagrams of $D^0\to\omega\gamma'$ (a) and $D^0\to\gamma\gamma'$ (b) through $cu\gamma'$ effective coupling in dimension-six operator.
        } 
	\label{fig:D02Gp}
\end{figure}

In this Letter, we focus on the first item of the dimension-six operator in Eq.~\eqref{eq:dimension-six operator}, which causes the $cu\gamma'$ coupling in the flavor-changing neutral current (FCNC) process of a charm quark with $j=1,k=2$.
In the SM, the FCNC processes are strongly suppressed by the Glashow-Iliopoulos-Maiani mechanism~\cite{Glashow:1970gm}, stating that these processes are forbidden at the tree level and can only happen through a loop diagram. The branching fraction (BF) of the charm FCNC process is expected to be smaller than $10^{-9}$ within the SM~\cite{Fajfer:2001sa,Paul:2011ar,Cappiello:2012vg,Li:2024moj}. But for the $cu\gamma'$ coupling, its FCNC originates from the NP energy scale, which is different from the SM. 
In the charm sector, the massless dark photon can be searched for in $D$ meson or $\Lambda^+_c$ baryon decays, such as $D\to V\gamma'$, $D\to\gamma\gamma'$, or $\Lambda^+_c\to p \gamma'$, where $V$ is a vector particle like $\rho$ or $\omega$. The BFs of these processes are in direct proportion to $|\mathbb{C}|^2+|\mathbb{C}_5|^2$~\cite{Su:2020yze}. Here $\mathbb{C}=\Lambda_{\rm{NP}}^{-2}(C^U_{12}+C^{U*}_{12})\nu/\sqrt{8}$ and $\mathbb{C}_5=\Lambda_{\rm{NP}}^{-2}(C^U_{12}-C^{U*}_{12})\nu/\sqrt{8}$ with the Higgs vacuum expectation value $\nu=246.2$ GeV~\cite{Plehn:2005nk}, which are determined by the NP energy scale $\Lambda_{\rm{NP}}$ and the up-type dimensionless coefficient $C^U_{12}$. From the constraint of the dark matter (DM) and the vacuum stability (VS) in the universe~\cite{Gabrielli:2016cut,Su:2020yze}, the allowed BF of the massless dark photon in charm FCNC processes can be enhanced to the order of $10^{-7}\sim10^{-5}$~\cite{Su:2020yze}. 
The sensitivity of the previous $\Lambda^+_c\to p \gamma'$~\cite{BESIII:2022vrr} search does still not reach the allowed region obtained from DM and VS~\cite{Gabrielli:2016cut,Su:2020yze,Li:2024iqv}. In this Letter, we search for the massless dark photon and probe the NP energy scale through the FCNC processes $D^{0}\to\omega\gamma'$ and $D^{0}\to\gamma\gamma'$ for the first time, which can be mediated via Feynman diagrams shown in FIG.~\ref{fig:D02Gp}, by analyzing $e^+e^-$ collision data of $7.9~\rm{fb^{-1}}$ at a center-of-mass energy of $\sqrt{s}=3.773$ GeV with the BESIII detector.

\vspace{-0.0cm}
\vspace{-0.0cm}

Details about the BESIII detector design and performance are provided elsewhere~\cite{BESIII:2023exq}. The simulated Monte Carlo (MC) samples, also described in Ref.~\cite{BESIII:2023exq}, are used to determine detection efficiencies and to estimate backgrounds. The generator of the signal MC samples is parameterized by the helicity amplitudes same as the radiative decay of the $D$ meson~\cite{Belle:2016mtj,BaBar:2008kjd}.
At $\sqrt{s}=3.773$ GeV, the $D^0\bar{D}^0$ meson pairs are produced from $\psi(3770)$ decays without accompanying hadrons, which provide an ideal opportunity to study invisible massless dark photon decays of $D$ mesons using the double tag (DT) method~\cite{MARK-III:1985hbd}. The $\bar{D}^0$ mesons are first tagged with the main hadronic-decay modes $\bar{D}^0\to K^+\pi^-$, $\bar{D}^0\to K^+\pi^-\pi^0$ and $\bar{D}^0\to K^+\pi^-\pi^+\pi^-$, and the selected candidates are referred to as the single tag (ST) sample. Here and throughout this letter, charge conjugations are always implied. Then, the signal processes $D^0\to\omega\gamma'$ and $D^0\to\gamma\gamma'$ are searched for in the system recoiling against the ST $\bar{D}^{0}$ meson, and the selected candidates are denoted as the DT sample. Here, $\omega$ is reconstructed through its decay $\omega\to\pi^+\pi^-\pi^0$, $\pi^0\to\gamma\gamma$, and $\gamma'$ is missing in the detector. The BFs of $D^0\to\omega\gamma'$ and $D^0\to\gamma\gamma'$ are calculated by
\begin{eqnarray}
\mathcal{B}(D^0\to\omega(\gamma) ~\gamma')= \frac{N_{\rm{sig}}/\hat{\epsilon}}{\mathcal{B}_{\rm{int}} \sum\limits_i N_{i}^{\rm{ST}}},
\label{eq:BF}
\end{eqnarray}
with $\sum\limits_i N_{i}^{\rm{ST}}=(6306.7\pm2.8)\times10^3$~\cite{BESIII:2023exq} and the effective efficiency
\begin{eqnarray}
\hat{\epsilon}=\sum_i \frac{\epsilon_i^{\rm{DT}}}{\epsilon_i^{\rm{ST}}} \times \frac{N_i^{\rm{ST}}}{\sum\limits_i N_i^{\rm{ST}}},
\end{eqnarray}
where $i$ indicates each mode of $\bar{D}^0\to\rm{hadrons}$, $N_{\rm{sig}}$ is the signal yield of the massless dark photon in data, $N_i^{\rm{ST}}$ is the ST yield of $\bar{D}^0$ meson samples in data, $\epsilon_i^{\rm{ST}}$ is the ST efficiency of $\bar{D}^0\to\rm{hadrons}$, $\epsilon_i^{\rm{DT}}$ is the DT efficiency of $\bar{D}^0\to\rm{hadrons}$, $D^0\to\omega(\gamma)\gamma'$, $\mathcal{B}_{\rm{int}}=\mathcal{B}(\omega\to\pi^+\pi^-\pi^0)\times\mathcal{B}(\pi^0\to\gamma\gamma)$ is obtained from Particle Data Group~\cite{pdg:2024} for $D^0\to \omega\gamma'$ and $\mathcal{B}_{\rm{int}}=1$ for $D^0\to \gamma\gamma'$.

    

The selection criteria of ST samples, the ST yield $N_i^{\rm{ST}}$, and the ST efficiency $\epsilon_i^{\rm{ST}}$ can be found in Ref.~\cite{BESIII:2023exq}. The selection criteria of $D^0\to\omega\gamma'$ and $D^0\to\gamma\gamma'$, based on the tagged $\bar{D}^0$ meson samples, are described below. To select $D^0\to\gamma\gamma'$, no additional charged track is allowed. 
The good charged track, particle identification (PID), and photon candidates are selected with the same strategy as outlined in Ref.~\cite{BESIII:2023exq}.
To select $D^0\to\omega\gamma'$, only events with exactly two selected charged tracks, both identified as pions with zero net charge, are retained for further analysis.
There should be at least one photon with energy larger than 0.5 GeV for $D^0\to\gamma\gamma'$ and at least two photons for $D^0\to\omega\gamma'$, where the two photons with minimum $\chi^2$ value of the kinematic fit~\cite{Yan:2010zze} constraining $M_{\gamma\gamma}$ to the nominal $\pi^0$ mass are regarded as the correct photons from the $\pi^0$ meson. To select the $\omega$ meson in the data samples, the invariant mass of the two photons $M_{\gamma\gamma}$ before the kinematic fit must be in the region of $[0.115,~0.150]\gevcc$, and the invariant mass $M_{\pi^+\pi^-\pi^0}$ of the $\omega$ candidate is required to be in the region of $[0.700,~0.850]\gevcc$. To further reduce the non-$\omega$ background, a kinematic fit~\cite{Yan:2010zze} constraining $M_{\gamma\gamma}$ to the nominal $\pi^0$ mass and $M_{\pi^+\pi^-\pi^0}$ to the nominal $\omega$ mass is performed to obtain the $\chi^2_{2\rm{C}}$ value which is required to be less than 44, optimized with the Punzi-optimization method~\cite{Punzi:2003bu}. To suppress the background with additional photons or $\pi^0$, the total energy of photon candidates other than those from the $\pi^0$ ($\gamma$) and the $\bar{D}^0$ meson ($E^{\rm{tot}}_{\rm{oth}.\gamma}$) is required to be less than 0.1 GeV for $D^0\to\omega\gamma'$ ($D^0\to\gamma\gamma'$). After these selections, there may still be some background particles flying to the endcap of the detector that cannot be effectively detected~\cite{Li:2024pox}, so the recoiling angle of $\bar{D}^0\omega$ ($\bar{D}^0\gamma$) is applied to veto these associated background events. The cosine of the recoiling angle is defined as $\cos\theta^{\rm{reccoil}}_{\bar{D}\omega(\gamma)}=\frac{|\vec{\textbf{p}}_{\rm{cms}} - \vec{\textbf{p}}_{\bar{D}^0} - \vec{\textbf{p}}_{\omega(\gamma)}|_{{z}}}{|\vec{\textbf{p}}_{\rm{cms}} - \vec{\textbf{p}}_{\bar{D}^0} - \vec{\textbf{p}}_{\omega(\gamma)}|}$, where $\vec{\textbf{p}}_{\rm{cms}}$ is the momentum of the center-of-mass in $e^+e^-$ collision, $\vec{\textbf{p}}_{\bar{D}^0}$ is the reconstructed momentum of the $\bar{D}^0$ meson, $\vec{\textbf{p}}_{\omega(\gamma)}$ is the reconstructed momentum of $\omega$ ($\gamma$), and the subscript $z$ refers to the $z$-component. To suppress these background events, a requirement of $|\cos\theta_{\bar{D}\omega(\gamma)}^{\rm{recoil}}|<0.7$ is applied. 
With the above selection criteria, 
the effective efficiency is estimated from the MC samples, which is $\hat{\epsilon}=(15.98\pm0.02)\%$ for $D^0\to\omega\gamma'$ and $\hat{\epsilon}=(52.18\pm0.05)\%$ for $D^0\to\gamma\gamma'$.
The main background after the selections comes from the $K^0_L$ associated background events, such as $D^0\to\omega K^0_L$ for $D^0\to\omega\gamma'$ and $D^0\to\pi^0 K^0_L$ for $D^0\to\gamma\gamma'$.

\vspace{-0.0cm}
\begin{figure}[htbp] \centering
	\setlength{\abovecaptionskip}{-1pt}
	\setlength{\belowcaptionskip}{10pt}

        {\includegraphics[width=0.5\textwidth]{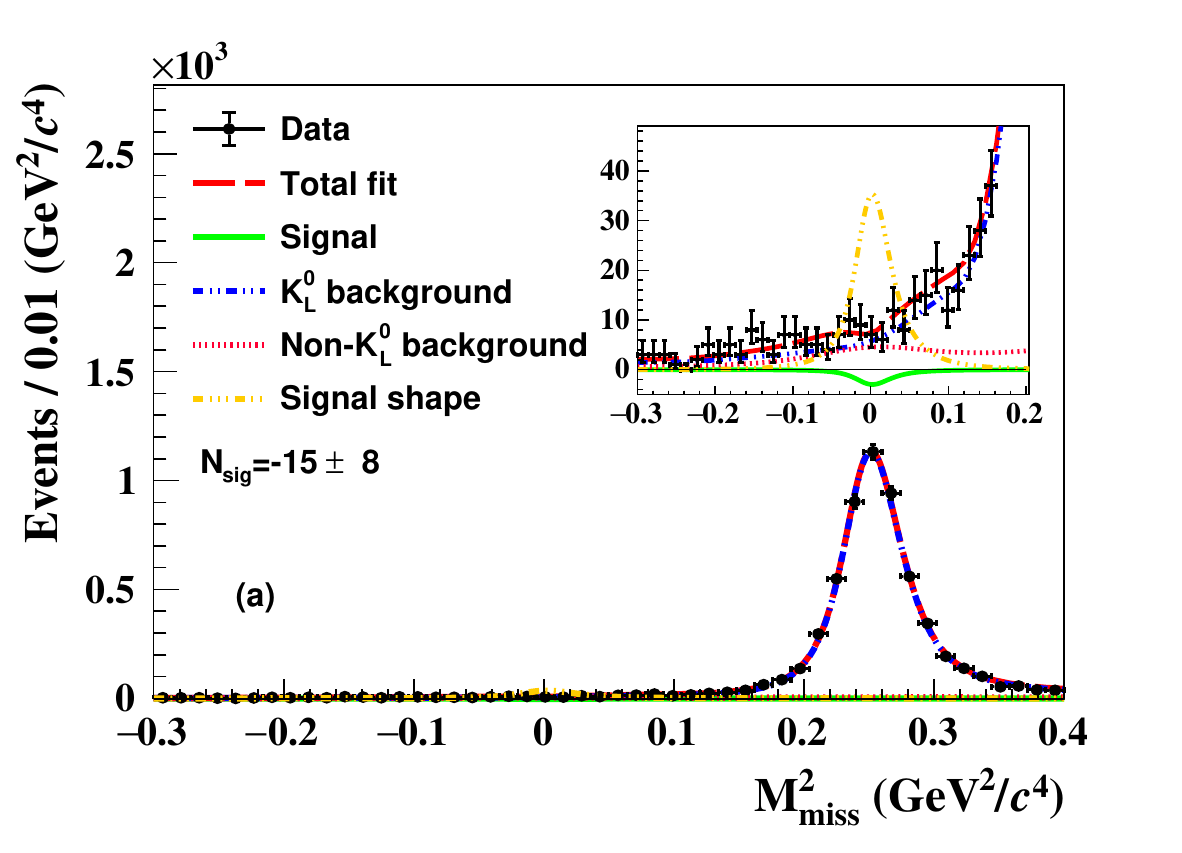}}\\
        {\includegraphics[width=0.5\textwidth]{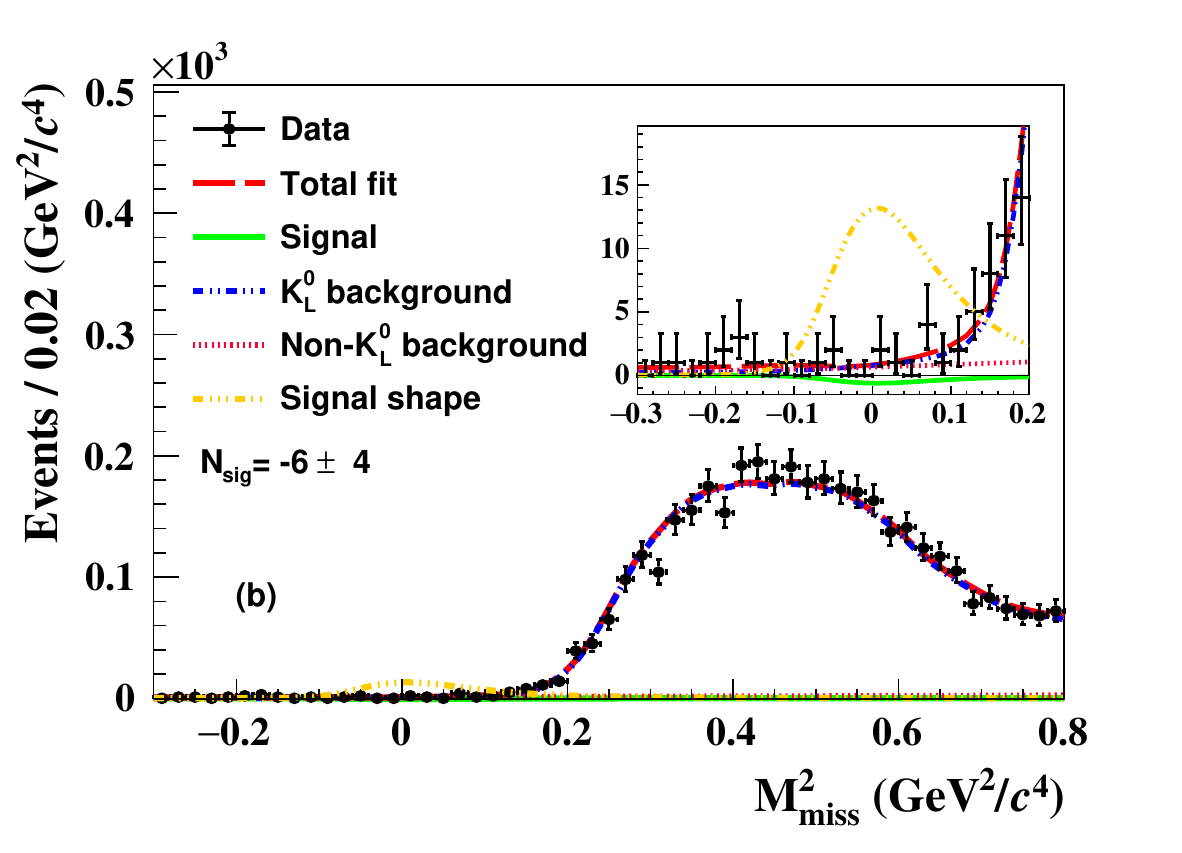}}\\
        
	\caption{The $M^2_{\rm miss}$ distributions of the accepted candidates of $D^0\to\omega\gamma'$ (a) and $D^0\to\gamma\gamma'$ (b). In the plot, the magnitude of the signal shape corresponds to $\mathcal{B}(D^0\to\omega\gamma')=2\times10^{-4}$ (a) and $\mathcal{B}(D^0\to\gamma\gamma')=4\times10^{-5}$ (b) respectively.
    } 
	\label{fig:fit}
\end{figure}
\vspace{-0.0cm}

The signals of the massless dark photon are extracted from an unbinned extended maximum likelihood fit on the distribution of the  square of the missing mass, $M^2_{\rm{miss}}$, defined as
\begin{eqnarray}
M^2_{\rm{miss}}=|p_{\rm{cms}} - p_{\bar{D}^0} - p_{\omega(\gamma)}|^2/\it{c}^{\rm{4}},
\label{eq:Mmiss2}
\end{eqnarray}
where $p_{\rm{cms}}$ is the four-momentum of the $e^+e^-$ center-of-mass system in the laboratory frame, $p_{\omega(\gamma)}$ is the kinematic fitted (reconstructed) four-momentum of $\omega(\gamma)$, $p_{\bar{D}^0}$ is the four-momentum of the $\bar{D}^0$ meson, achieved by the kinematic fit~\cite{Yan:2010zze} constraining $M_{\gamma\gamma}$ to the nominal $\pi^0$ mass and $M_{K^+\pi^-}$, $M_{K^+\pi^-\pi^0}$, $M_{K^+\pi^-\pi^+\pi^-}$ to the nominal $\bar{D}^0$ meson mass. In the fit, the background is separated into the $K^0_{L}$-related and non-$K^0_{L}$ backgrounds~\cite{Zhou:2020ksj}. 
The background shape is derived from the inclusive MC sample~\cite{BRUN199781}, with the number of non-$K^0_{L}$ background events assumed to follow a Gaussian distribution and constrained by the MC simulation (referred to as the Gaussian constraint), while the number of $K^0_{L}$-related background events is left as a floating parameter.
The signal is derived by the simulated shape convolved with a Gaussian function $G(\mu,\sigma)$, where $\mu$ and $\sigma$ are restrained to the values obtained from the control samples $D^0\to\omega K^0_S$ ($D^0\to\pi^0 K^0_S$) for $D^0\to\omega\gamma'$ ($D^0\to\gamma\gamma'$). The fit results are shown in FIG.~\ref{fig:fit}, with the massless dark photon signal yield $N_{\rm{sig}}=-15\pm8$ for $D^0\to\omega\gamma'$ and $N_{\rm{sig}}=-6\pm4$ for $D^0\to\gamma\gamma'$.


        

The systematic uncertainty sources for the BF measurement include ST yield, intermediate BF, signal generator, DT signal efficiency, and signal extraction. With the DT method, several systematic uncertainties associated with the ST selection can be canceled without impacting the BF measurement. The uncertainty of ST yield is assigned as 0.1\%~\cite{BESIII:2023exq}. 
The uncertainty of the generator is estimated from the efficiency difference compared with a flat angular generation of $\gamma'$ (phase space model), which is 1.3\% (0.6\%) for $D^0\to\omega\gamma'$ ($D^0\to\gamma\gamma'$).
The uncertainty of the BF of $\omega\to\pi^+\pi^-\pi^0$ is 0.8\% and that of $\pi^0\to\gamma\gamma$ is negligible~\cite{pdg:2024}. 
The uncertainty of photon detection is assigned as 1.0\% per photon~\cite{BESIII:2015kvk}. The uncertainties of pion tracking and PID are studied from the control sample $J/\psi\to\pi^+\pi^-\pi^0$, which is assigned as 0.9\% for tracking and 1.1\% for PID of the two pions.
The uncertainties of other selections are estimated from the control sample $D^0\to\omega K^0_S$ ($D^0\to\pi^0 K^0_S$) for $D^0\to\omega\gamma'$ ($D^0\to\gamma\gamma'$), where the $K^0_S$ meson is regarded as a missing particle. For $D^0\to\omega \gamma'$, the uncertainty is 0.2\% for the $M_{\gamma\gamma}$ selection, negligible for the $M_{\pi^+\pi^-\pi^0}$ selection, 3.1\% for the $\chi^2_{2\rm{C}}$ selection, 5.9\% for the $E^{\rm{tot}}_{\rm{oth}.\gamma}$ selection and 1.1\% for the $\cos\theta^{\rm{recoil}}_{\bar{D}\omega}$ selection, respectively. For $D^0\to\gamma \gamma'$,  the uncertainty is 2.4\% for the $E^{\rm{tot}}_{\rm{oth}.\gamma}$ selection and is 0.6\% for the $\cos\theta^{\rm{recoil}}_{\bar{D}\gamma}$ selection, respectively. The total systematic uncertainty is calculated by summing up all sources in quadrature, yielding 7.4\% for $D^0\to\omega\gamma'$ and 2.7\% for $D^0\to\gamma\gamma'$, as shown in TABLE~\ref{tab:sys}.
For the uncertainty from the signal extraction, the signal shape is convolved with a Gaussian function to describe the difference where the parameter of the Gaussian function is in a Gaussian constraint within its uncertainty, the $K^0_L$ background yield is floating in the fit, and the non-$K^0_L$ background yield is also floating in a Gaussian constraint within its uncertainty. The uncertainty of signal extraction is negligible.

\begin{table}[!htbp]
\caption{Summary of the systematic uncertainties for the measurement of the BF of $D^0\to\omega\gamma'$ and $D^0\to\gamma\gamma'$. The total value is calculated by summing up all sources in quadrature.}
\setlength{\abovecaptionskip}{1.2cm}
\setlength{\belowcaptionskip}{0.2cm}
\begin{center} 
\footnotesize
\vspace{-0.0cm}
\begin{tabular}{l|cc}
\hline \hline
            \specialrule{0em}{2pt}{2pt}
		\multirow{2}{*}{Sources} & \multicolumn{2}{c}{Related uncertainties ($\%$)}\\\specialrule{0em}{2pt}{2pt}
                                      & $D^0\to\omega\gamma'$ & $D^0\to\gamma\gamma'$ \\\specialrule{0em}{2pt}{2pt}
            \hline\specialrule{0em}{2pt}{2pt}
            Tracking & 0.9 & $-$ \\\specialrule{0em}{2pt}{2pt}
            PID & 1.1 & $-$ \\\specialrule{0em}{2pt}{2pt}
            Photon detection & 2.0 & 1.0 \\\specialrule{0em}{2pt}{2pt}
            $M_{\gamma\gamma}$ & 0.2 & $-$ \\\specialrule{0em}{2pt}{2pt}
            $M_{\pi^+\pi^-\pi^0}$ & negligible & $-$ \\\specialrule{0em}{2pt}{2pt}
            $\chi^2_{2\rm{C}}$ & 3.1 & $-$ \\\specialrule{0em}{2pt}{2pt}
            $E^{\rm{tot}}_{\rm{oth}.\gamma}$ & 5.9 & 2.4 \\\specialrule{0em}{2pt}{2pt}
            $\cos\theta^{\rm{reccoil}}_{\bar{D}\omega}$ & 1.1 & $-$ \\\specialrule{0em}{2pt}{2pt}
            $\cos\theta^{\rm{reccoil}}_{\bar{D}\gamma}$ & $-$ & 0.6 \\\specialrule{0em}{2pt}{2pt}
            Single tag yield & 0.1 & 0.1 \\\specialrule{0em}{2pt}{2pt}
            $\mathcal{B}(\omega\to\pi^+\pi^-\pi^0)$ & 0.8 & $-$ \\\specialrule{0em}{2pt}{2pt}
            $\mathcal{B}(\pi^0\to\gamma\gamma)$ & negligible & $-$ \\\specialrule{0em}{2pt}{2pt}
            Signal generator & 1.3 & 0.6 \\\specialrule{0em}{2pt}{2pt}
            \hline\specialrule{0em}{2pt}{2pt}
            Total & $7.4$ & $2.7$ \\\specialrule{0em}{2pt}{2pt}

\hline \hline
\end{tabular}
\label{tab:sys}
\end{center}
\end{table}
\vspace{-0.0cm}

Since no significant excess of signal above the background is observed, a UL on the BF is set using a Bayesian approach following Ref.~\cite{BESIII:2023fqz}, where the BF is calculated by Eq.~\eqref{eq:BF} and the systematic uncertainty is estimated with the method in Refs.~\cite{Liu:2015uha,Stenson:2006gwf}. The UL on the BF at the 90\% CL is calculated by integrating the likelihood distribution with different signal assumptions to the $90\%$ region, which is $\mathcal{B}(D^0\to\omega\gamma')<1.1\times10^{-5}$ and $\mathcal{B}(D^0\to\gamma\gamma')<2.0\times10^{-6}$. 
Note that in the $D^0\to\omega\gamma'$ measurement, the non-$\omega$ contribution can not be fully removed, and the current UL of $\mathcal{B}(D^0\to\omega\gamma')$ is a conservative estimation.

\begin{figure}[htbp] \centering
	\setlength{\abovecaptionskip}{-1pt}
	\setlength{\belowcaptionskip}{10pt}

        {\includegraphics[width=0.5\textwidth]{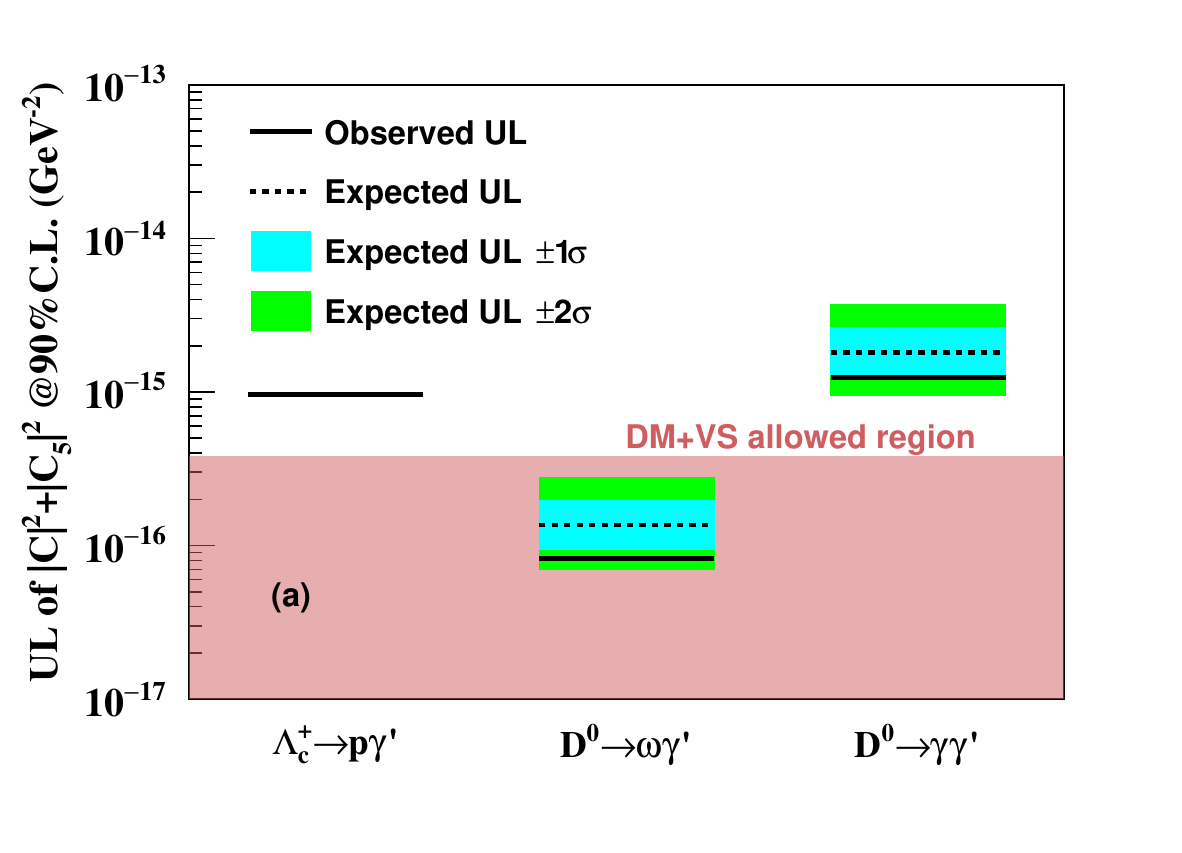}}\\
        {\includegraphics[width=0.5\textwidth]{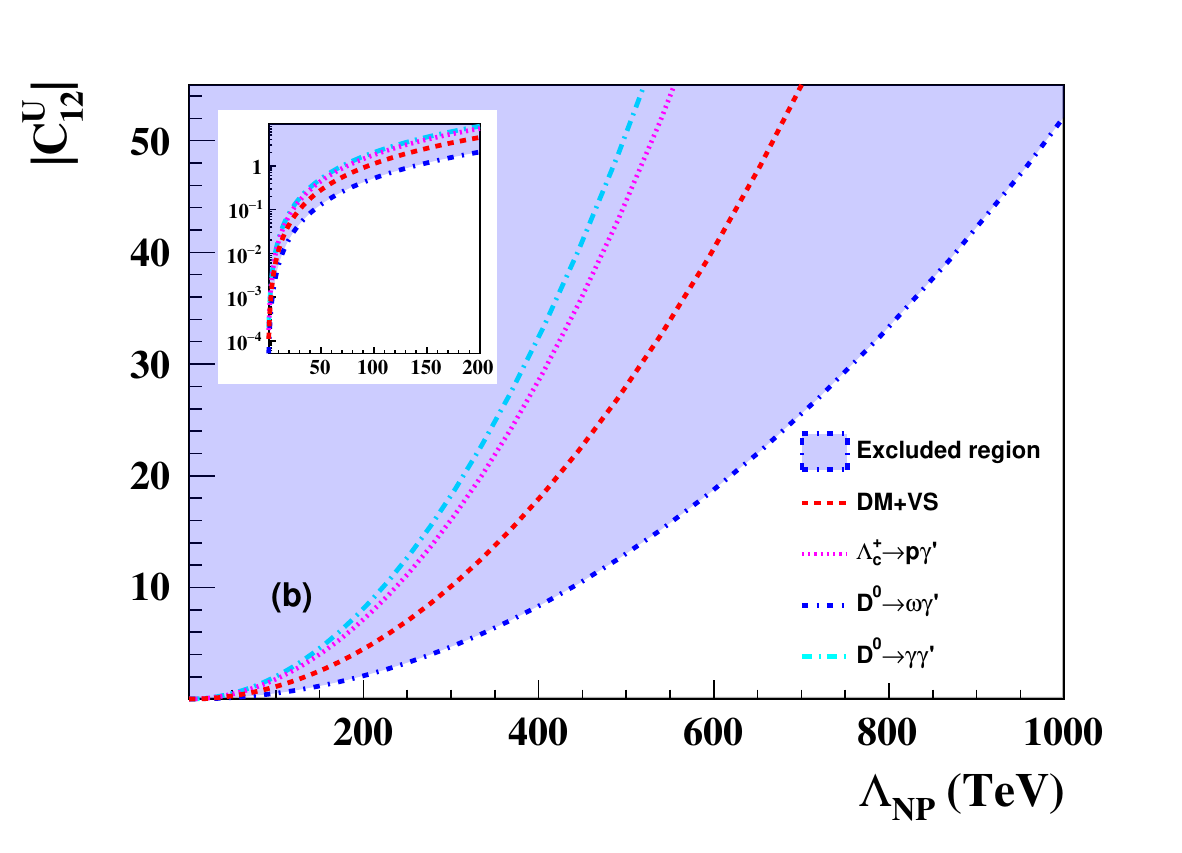}}\\
        
	\caption{(a) The ULs of $|\mathbb{C}|^2+|\mathbb{C}_5|^2$ for $D^0\to\omega\gamma'$ and $D^0\to\gamma\gamma'$ obtained in this work and $\Lambda^+_c\to p\gamma^\prime$ in Ref.~\cite{BESIII:2022vrr}. 
         (b) The two-dimensional constraints on the NP energy scale $\Lambda_{\rm{NP}}$ and the dimensionless coefficient $|C^{U}_{12}|$. 
        } 
	\label{fig:UL}
\end{figure}

The operator Eq.~(\ref{eq:dimension-six operator}) may cover some new dark-sector particles with very heavy mass in the NP energy scale $\Lambda_{\rm{NP}}$~\cite{Gabrielli:2013jka,Gabrielli:2016cut}. Up to now, no new particles have been found up to the mass of $\sim 1$ TeV, but some anomalies require an energy scale above the electroweak energy scale $\Lambda_{\rm{EW}}$ ($\sim 100$ GeV).
Similar to the $\beta$ decay observed in low energy experiments~\cite{Chadwick:1914zz}, where a missing neutrino within the four-fermion effective coupling~\cite{Fermi:1934hr} can predict the electroweak energy scale at about 100 GeV~\cite{UA1:1983crd}, the massless dark photon could provide a portal for exploring $\Lambda_{\rm{NP}}$ beyond the TeV magnitude.
Since the BF of massless dark photon production is related to $|\mathbb{C}|^2+|\mathbb{C}_5|^2$ and directly includes the NP energy scale~\cite{Su:2020yze}, the constraint on $|\mathbb{C}|^2+|\mathbb{C}_5|^2$ can be performed as well.
The UL of $|\mathbb{C}|^2+|\mathbb{C}_5|^2$ is shown in FIG.~\ref{fig:UL}. For $D^0\to\omega\gamma'$, one sees that $|\mathbb{C}|^2+|\mathbb{C}_5|^2<8.2\times10^{-17}\gev^{-2}$, reaching the DM and VS allowed region~\cite{Gabrielli:2016cut,Su:2020yze} for the first time and surpassing the previous $\Lambda^+_c\to p\gamma'$ measurement~\cite{BESIII:2022vrr,Li:2024iqv} by more than an order of magnitude.
The channel $D^0\to\gamma\gamma'$ has a better UL of the BF but a worse constraint on $cu\gamma'$ coupling due to an additional electromagnetic vertex in FIG~\ref{fig:D02Gp}(b).
The two-dimensional constraint on the NP energy scale $\Lambda_{\rm{NP}}$ and the up-type dimensionless coefficient $C^U_{12}$ is given in FIG~\ref{fig:UL} (b). Our result currently resemble the best constraint on the NP parameter space. Assuming $|C^U_{12}|=1$, our results can exclude NP energy scales below 138 TeV in the dark sector, which is approximately ten times the energy reached at the Large Hadron Collider~\cite{Evans2008LHCM}, suggesting a challenge of directly detecting superheavy particles at the NP energy scale associated with the $cu\gamma'$ coupling in the present collider settings. Note that the value of $|C^U_{12}|$ is model-dependent.

\vspace{-0.0cm}
\vspace{-0.0cm}

In summary, we search for the massless dark photon and place constraints on the new physics scale in the charm FCNC processes $D^0\to\omega\gamma'$ and $D^0\to\gamma\gamma'$ for the first time. 
Using $7.9~\rm{fb}^{-1}$ of $e^+e^-$ collision data at $\sqrt{s}=3.773$ GeV, no significant signals are observed. The ULs on the BFs are set to be $\mathcal{B}(D^0\to\omega\gamma')<1.1\times10^{-5}$ and $\mathcal{B}(D^0\to\gamma\gamma')<2.0\times10^{-6}$ at the 90\% confidence level.
The constraint on the parameter related to the new physics energy scale is found to be $|\mathbb{C}|^2+|\mathbb{C}_5|^2<8.2\times10^{-17}\gev^{-2}$ at the 90\% confidence level, with an improvement over the previous results by one order of magnitude and exploring the DM and VS allowed space for the first time.
Our study provides a new stringent result for the massless dark photon, an important component within the dark photon framework in the dark sector. However, in comparison to the massive dark photon, researches on the massless counterpart remain limited on the experimental side. Future studies of the $sd\gamma'$ coupling in $K_L\to\gamma\gamma'$ at KOTO~\cite{10.1093/ptep/pts057}, the $bs\gamma'$ coupling in $B^+\to K^{*+}\gamma'$ at BELLE-II~\cite{Belle-II:2010dht} or in $B_s \to \phi\gamma'$ at LHCb~\cite{LHCb:2008vvz,LHCb:2014set}, and the $tc\gamma'$ coupling in $t\to c\gamma'$ at CMS~\cite{CMS:2008xjf} or ATLAS~\cite{ATLAS:2008xda}, can provide more comprehensive constraints on the massless dark photon.

\textbf{Acknowledgement}

The BESIII Collaboration thanks the staff of BEPCII and the IHEP computing center for their strong support. This work is supported in part by National Key R\&D Program of China under Contracts Nos. 2023YFA1606000, 2020YFA0406400, 2020YFA0406300; National Natural Science Foundation of China (NSFC) under Contracts Nos. 11635010, 11735014, 11935015, 11935016, 11935018, 12025502, 12035009, 12035013, 12061131003, 12175321, 12192260, 12192261, 12192262, 12192263, 12192264, 12192265, 12221005, 12225509, 12235017, 12361141819; the Chinese Academy of Sciences (CAS) Large-Scale Scientific Facility Program; the CAS Center for Excellence in Particle Physics (CCEPP); Joint Large-Scale Scientific Facility Funds of the NSFC and CAS under Contract No. U1832207, U1932101; 100 Talents Program of CAS; The Institute of Nuclear and Particle Physics (INPAC) and Shanghai Key Laboratory for Particle Physics and Cosmology; German Research Foundation DFG under Contracts Nos. FOR5327, GRK 2149; Istituto Nazionale di Fisica Nucleare, Italy; Knut and Alice Wallenberg Foundation under Contracts Nos. 2021.0174, 2021.0299; Ministry of Development of Turkey under Contract No. DPT2006K-120470; National Research Foundation of Korea under Contract No. NRF-2022R1A2C1092335; National Science and Technology fund of Mongolia; National Science Research and Innovation Fund (NSRF) via the Program Management Unit for Human Resources \& Institutional Development, Research and Innovation of Thailand under Contracts Nos. B16F640076, B50G670107; Polish National Science Centre under Contract No. 2019/35/O/ST2/02907; Swedish Research Council under Contract No. 2019.04595; The Swedish Foundation for International Cooperation in Research and Higher Education under Contract No. CH2018-7756; U. S. Department of Energy under Contract No. DE-FG02-05ER41374.

\bibliographystyle{apsrev4-1}
\bibliography{mybib.bib}

\end{document}